\documentclass[prd,aps,twocolumn,showpacs,showkeys]{revtex4}
\input epsf
\begin{document}

\title{Attractive Casimir effect in an infrared modified gluon bag model}

\author{L. E. Oxman$^a$, N. F. Svaiter$^b$ and R. L. P. G. Amaral$^a$\\~}
\affiliation{{$^a$ Instituto de F\'{\i}sica, Universidade Federal Fluminense}\\
{Av. Litor\^anea S/N, Boa Viagem,}\\
{Niter\'oi, RJ 24210-340, Brazil.}\\ \\
{$^b$Centro Brasileiro de Pesquisas F\'{\i}sicas - CBPF}\\
{Rua Dr. Xavier Sigaud 150, Rio de Janeiro, RJ 22290-180, Brazil.}}
\date{\today}

\begin{abstract}
In this work, we are motivated by previous attempts to derive the
vacuum contribution to the bag energy in terms of familiar Casimir
energy calculations for spherical geometries. A simple infrared modified model
is introduced which allows studying the effects of the analytic structure
as well as the geometry in a clear manner. In this context, we show that if a
class of infrared vanishing effective gluon propagators is considered,
then the renormalized vacuum energy for a spherical bag is attractive,
as required by the bag model to adjust hadron spectroscopy.
\end{abstract}

\maketitle
%%%%%%%%%%%%%%%%%%%%%%%%%%%%%%%%%%%%%%%%
\section{Introduction}
%%%%%%%%%%%%%%%%%%%%%%%%%%%%%%%%%%%%%%%

In the MIT bag model, a hadron is regarded as a finite region of
space where quark and gluon fields are confined. The bag energy
necessary to fit hadron masses must contain a term of the form
$-Z/R$, where $R$ is the bag radius and $Z$ is a phenomenological
positive constant of order one \cite{dejjk}. It is commonly
accepted that this term is essentially a manifestation of zero
point energies of the confined fields, however an understanding of
this effect in the context of Casimir energy calculations remains
intriguing.

These calculations consider gluons as free massless particles inside
the bag, an assumption based on asymptotic freedom. Boyer, Davies,
Bender and Hays \cite{boyer,davies, bender}, and many others,
studied massless scalar, vector and fermionic fields assuming they
are confined in a spherical region of space. A systematic study
has been developed by Milton \cite{milton1,milton2,milton3}. In all
these studies, the renormalized zero-point energy contribution to
the MIT bag energy contains the term $\sim 1/R$, but with a
negative $Z$ ($\approx -0.7$), that is, a repulsive effect is
obtained, instead of the attractive one necessary to adjust
hadron spectroscopy. To the best of our knowledge, there is no
explanation in the literature for this discrepancy. For a
discussion of the bag model and the current status of Casimir
calculations in the bag see refs. \cite{Don} and \cite{Milton},
respectively.

A resolution to the sign problem is hard to accomplish as in
general there is no intuition whether the Casimir energy should be
positive or negative in a given situation. In general, the sign of
the Casimir energy may depend on spacetime dimensionality, the
type of boundary conditions, the shape of the boundaries, the spin
of the field, etc. For a discussion of these issues, see for
example refs. \cite{wolfram} \cite{nami2} \cite{nami3}.

On the other hand, any treatment of QCD with boundaries requires
simplifying assumptions
to render the calculation possible. In this regard, we note that the above
mentioned simple model considering free massless gluons inside the bag only
encodes nonperturbative information
by imposing boundary conditions on the bag boundary, while the free
propagator
is taken over the whole range of momenta that can be accommodated inside
the bag.
This means, momenta ranging typically from the inverse bag radius
up to the ultraviolet region. We would like to stress that infrared modes
are more sensitive to the bag geometry, and these are precisely modes in
the nonperturbative region of QCD. This situation, shows that any
improvement to obtain an attractive Casimir energy must take into account
intrinsic nonperturbative effects inside the bag such as the analytic structure QCD.

At present, a derivation of the Casimir energy from first principles seems 
hopelessly out of reach. 
In this work we would like to propose instead a possible understanding
of the attractive nature of vacuum fluctuations, based on a simple bag model 
extended to incorporate an infrared modified gluon propagator. The associated 
infrared behavior will be motivated by recent studies on the analytic
structure of the confined gluon propagator in pure QCD
\cite{Alkofer}.

\section{Simple infrared modified bag model}
\label{irmodel}
%%%%%%%%%%%%%%%%%%%%%%%%%%%%%%%%%%%%%%%%%%%%%%%%%%%%%%%%%%%%%%%%%%%%%%%%

In QCD it is believed that colored objects such as quarks and gluons cannot exist
in asymptotic states. They should give place instead to the hadronic spectrum. These open problems are associated with a nonperturbative regime, and suppossed to be driven by the dynamical generation of a mass scale, which separates the perturbative and nonperturbative phases.  

In the bag model, the boundary condition is devised to represent the nonperturbative physics
associated with the finite size of colorless objects such as hadrons.
 
Now, regarding the renormalized vacuum energy, if it is computed by considering the gluon sector only (quenched model), the inclusion of additional nonperturbative information is a sensible modification to be considered. 

In this regard, note that such a calculation would be based on pure QCD proved by means of a boundary condition on a sphere of radius R. Of course, for very small $R$, the effect of the boundaries would  be relevant, while, because of asymptotic freedom, the physics inside the bag would essentially be perturbative. On the other hand, for very large $R$, the model would behave as pure QCD without boundaries, displaying the intrinsic nonperturbative behavior of the gluon sector.

For typical values of $R$ associated with nonperturbative objects such as hadrons, both
effects should be relevant when computing the renormalized vacuum energy: finite size  effects representing the hadron boundaries, as well as intrinsic nonperturbative behavior in the gluon sector. In particular, possible nonperturbative analytic properties in this sector constitute an interesting aspect to be explored. 

In this section, according to the discussion above, an infrared modified bag model is presented. The basic physical inputs are the following:\\

$\bullet$ As usual, we will consider the
hadron as a finite region in space where the fields are confined by
imposing a boundary condition at the bag boundary.\\

$\bullet$ Additional intrinsic nonperturbative information will be
parametrized in a model gluon propagator $G(k^2)$ associated with an
effective quadratic gluon action inside the bag. Although this procedure
cannot be justified from first principles, similar ideas have been already
used in different hadron models where some interesting results have been
obtained \cite{Gastao}.
This enables the computation of the vacuum energy by means of the usual
formula,
\begin{eqnarray} E_C&=&\frac{i}{2T}
{\rm tr}\,{\ln{G}}\nonumber \\
&=&\frac{i}{2}\sum_n \int \frac{dk_0}{2\pi}\ln{G (k_0,k_n)},
\label{cas}
\end{eqnarray}
with $G(k^2)$ in the place of the free gluon propagator
$1/(k^2+i\epsilon)$. In this equation,
$T$ is the infinite time that the configuration exists and $k_n^2$ refers
to the eigenvalues of the Laplace operator
in the corresponding geometrical configuration. As in former studies, we
are considering a scalar Casimir energy
to be multiplied by the number of polarization modes of the gluon field
(including color) (see ref. \cite{Milton}).\\

$\bullet$ The model gluon propagator $G(k^2)$ incorporates nonperturbative
effects such as pole suppression. That is, we will change the free
$1/(k^2+i\epsilon)$ ($k^2=k_0^2-
{\bf k}^2$) propagator used
in previous works, which is valid in the ultraviolet region due to
asymptotic freedom, by an infrared modified one. After a Wick rotation $k^2+i\epsilon 
\to -\bar{k}^2$, the behavior of the modified propagator can be defined in euclidean 
$\bar{k}$ momentum space,
\begin{equation}G(-\bar{k}^2)\sim \left\{
\begin{array}{ll}
{\cal R}(\bar{k}^2)\, (1/\bar{k}^2), &  \bar{k}^2~ {\rm large} \\
{\cal R}(0)\, (\bar{k}^2)^\lambda, ~~\lambda>0,  &  \bar{k}^2~ {\rm small}.
\end{array}\right.
\label{beh}
\end{equation}
This type of infrared vanishing behavior has been
obtained in different scenarios (see the discussion below). For large
$\bar{k}^2$, we have also included a possible ${\cal R}(\bar{k}^2)$ factor
encoding running coupling constant information. Note also that the free
propagator would correspond to $\lambda=-1$ and ${\cal R}\equiv 1$. \\

When computing Casimir energies, we believe that the above
mentioned simple model can capture the essential modifications
implied by the analytic structure of QCD. With regard to the
behavior in (\ref{beh}), we are motivated by recent progress on
the form of the gluon propagator in pure QCD. In general, because
of confinement, we know that the gluon propagator must suffer a
dramatic change when $k^2\to 0$: as gluons cannot appear in
asymptotic states, the free pole at $k^2=0$ must
disappear in the complete theory. In fact, many authors coincide
that the exact propagator should be infrared finite when $k^2\to
0$. Indeed, an infrared vanishing behavior has been obtained in
different scenarios: by studying the Schwinger-Dyson equation
\cite{Alkofer04}, by restricting the path integral so as to avoid
Gribov copies \cite{Gribov}, and also in a Lattice
formulation \cite{Zwanziger}. In other words, different scenarios
point to an intrinsic analytic structure in pure QCD associated with 
an infrared suppressed propagator, describing the so called 
``confined'' gluons, in contrast to an infrared enhanced ``confining'' 
$1/k^4$ behavior (for a review, see ref. \cite{Alkofer}).

In particular, studies of the (Landau gauge) Schwinger-Dyson equation in pure QCD
 lead to an infrared vanishing 
gluon propagator $G(k^2)(\eta^{\mu \nu}-k^\mu k^\nu/k^2)$, with
the form  \cite{Alkofer04},
\begin{equation}
G(k^2)=(-k^2-i\epsilon)^{\lambda}(-k^2+\Lambda^2-i\epsilon)^{-(\lambda+1)}{\cal R}(-k^2-i\epsilon),
\label{SD}
\end{equation}
where $\lambda>0$ and $\Lambda$ is of the order of
$\Lambda_{QCD}$. Note that after a Wick rotation, $-k^2-i\epsilon
\to \bar{k}^2$, this ansatz satisfies the infrared vanishing
behavior given in (\ref{beh}). In the Schwinger-Dyson context, the
additional condition $\lambda < 0.4$ is satisfied, and the factor
${\cal R}(\bar{k}^2)$ represents the perturbative running. That is, for
$\bar{k}^2>>\Lambda^2$,
\begin{equation}
{\cal R}(\bar{k}^2)=\alpha^{-\gamma}
\makebox[.5in]{,}
\alpha \sim  \frac{{\rm const.}}{\ln (\bar{k}^2/\Lambda^2)}
\makebox[.5in]{,}
-\gamma> 0.
\label{Rfact}
\end{equation}

On the other hand, the Gribov mechanism is based on the restriction of the
path integral domain so as to avoid copies of the gauge fields in the pure
gluon theory defined on euclidean space.
This is necessary as in the nonperturbative region the usual gauge
conditions do not fix the gauge completely. In Landau gauge, the
associated nonperturbative gluon propagator is given by,
\begin{equation}\label{Schwinger-D}
G_{{\rm Gribov}}(-\bar{k}^2)=\frac{\bar{k}^2}{\bar{k}^4+M^4},
\label{G}
\end{equation}
which satisfies the behavior (\ref{beh}) with $\lambda=1$ and ${\cal R}\equiv 1$.

As we are concerned with a model gluon propagator, we will
first develop a Casimir energy calculation based on the associated general
analytic
properties displayed in Minkowski space, not relying on a specific form for
the propagator. We will obtain a representation for the Casimir energy
where the analytic structure of the model and the effects of geometry will
be clearly distinguished.
Only at the end we will discuss the effect of pole suppression on the
renormalized vacuum energy. For this aim, we will consider for instance a
model gluon propagator of the
form given in eq. (\ref{SD}), which displays a well defined analytic
structure in Minkowski
space. In this regard, note that the Gribov ansatz cannot be rotated back
to Minkowski space because of the complex poles in eq. (\ref{G}); as is
well known, the Gribov scenario is only defined in euclidean space.

\section{Casimir energy and the analytic structure of G}
%%%%%%%%%%%%%%%%%%%%%%%%%%%%%%%%%%%%%%%%%%%%%%%%%%%%%%%%%%%%%%%%%%%%%%%%
\label{casana}

In order to study the effect implied by the analytic structure of the modified propagator,
we will start by representing the logarithmic derivative of $G$ in terms 
of the decomposition,
\begin{equation} \frac{d}{{d\cal A}}\ln G ({\cal A}+i\epsilon)= \int_0^\infty
d\mu^2\beta(\mu^2)\frac{1}{-{\cal A} +\mu^2-i\epsilon},
\label{dec}
\end{equation}
where ${\cal A}=k^2$ and $\beta$ is given by,
\begin{equation} \beta(k^2)=\frac{1}{2\pi i}\left(\frac{d}{d {\cal A}}\ln
G({\cal A}+i\epsilon) -\frac{d}{d{\cal A}}  \ln G({\cal A}-i\epsilon)\right).
\end{equation}
Defining the jump at the discontinuity of an analytic function $F({\cal A})$,
\begin{equation}
\delta [F({\cal A})]\equiv F({\cal A}+i\epsilon)-F({\cal A}-i\epsilon),
\label{definit}
\end{equation}
we can also write,
\begin{equation}
\beta(k^2)=\frac{1}{2\pi i}
\delta \left[ \frac{d}{d{\cal
A}}\ln G({\cal A})\right].
\label{discontinuity}
\end{equation}
In this manner, from eq. (\ref{dec}), we have the representation,
\begin{equation} 
\ln G ({\cal A}+i\epsilon)= C +\int_0^\infty
d\mu^2\beta(\mu^2)\ln G_{\mu^2}({\cal A}+i\epsilon),
\label{simrep}
\end{equation}
with $C={\rm const.}$ and $G_{\mu^2}({\cal A})=(-{\cal A}+\mu^2)^{-1}$ being the free propagator for a field with mass parameter $\mu^2$.

In fact, when defining the Casimir energy in eq. (\ref{cas}), a parameter $\chi^2$ with the dimension of $[{\rm mass}]^2$ multypling $G$ must be introduced, so as to have a dimensionless argument in the logarithm. In an equivalent manner, we can introduce 
a parameter $\xi^2$, $[\xi]={\rm mass}$, multiplying $G_{\mu^2}$ in the second member 
of eq. (\ref{simrep}), and absorb the constant $C$ in its definition; then, we have,
\begin{equation} 
\ln \left( \chi^2 G \right)= \int_0^\infty
d\mu^2\beta(\mu^2)\ln \left( \xi^2 G_{\mu^2} \right).
\label{simrepad}
\end{equation}
Now, considering this replacement in eq. (\ref{cas}), we can take the formal trace in the second member of eq. (\ref{simrepad}), to obtain the Casimir energy in terms of the regularized representation,
\begin{eqnarray}
E_C&=& \int d\mu^2 \beta(\mu^2) 
\frac{i}{2} \left. \sum_n \int \frac{dk_0}{2\pi} \ln{\left( \xi^2 G_{\mu^2} \right)}\right|_{\rm reg}.\nonumber \\
\label{casreg}
\end{eqnarray}
In other words we can write,
\begin{equation}
E_C=\int d\mu^2 \beta(\mu^2) E_C(\mu^2),
\label{summodes2}
\end{equation}
where
\begin{equation}
E_C(\mu^2)=\frac{i}{2} \left. \sum_n \int \frac{dk_0}{2\pi} \ln{\left( \xi^2 G_{\mu^2}({\cal A}_n+i\epsilon) \right)}\right|_{\rm reg}
\label{Eeff}
\end{equation}
is the Casimir energy for free field modes with mass parameter $\mu^2$, obtained by means of the regularized partition function (${\cal A}_n=k_0^2-k_n^2$). After a Wick rotation $k_0 \rightarrow i\bar{k}_0$, we can also write,
\begin{equation}
E_C(\mu^2)=\frac{1}{2} \left. \sum_n \int \frac{d\bar{k}_0}{2\pi} \ln{\left( 
\frac{\bar{k}_0^2+k_n^2+\mu^2}{\xi^2} \right)}\right|_{\rm reg},
\label{Eeffeuc}
\end{equation}
which can be defined, for instance, by using the zeta function regularization technique, 
\begin{equation}
E_C(\mu^2)=-\frac{1}{2}\left.\frac{d~}{ds} \zeta_4(s)\right|_{s=0},
\label{zetaeff}
\end{equation}
\begin{equation}
\zeta_4(s)=\sum_n \int \frac{d\bar{k}_0}{2\pi} 
\left(\frac{\bar{k}_0^2+k_n^2+\mu^2}{\xi^2} \right)^{-s}.
\end{equation}

As discussed in ref. \cite{wipf}, the zeta function result for $E_C(\mu^2)$ in eq. (\ref{zetaeff}) differs in general from the Casimir energy obtained by the zeta function regularization of the summation of the free field energy modes, $E_{\rm mode}=(1/2)\sum_n (k_n^2+\mu^2)^{1/2}|_{reg}$,
\begin{equation}
E_{\rm mode}(\mu^2)=\frac{\xi}{2} \lim_{\epsilon\to 0} \frac{1}{2}\left[\zeta_3(-1/2+\epsilon)+ \zeta_3(-1/2-\epsilon)\right],
\label{modezeta}
\end{equation}
\begin{equation}
\zeta_3(s)=\sum_n \left(\frac{k_n^2+\mu^2}{\xi^2} \right)^{-s}.
\label{zetatres}
\end{equation}
In that reference, the following relationship has been established,
\begin{equation}
E_C(\mu^2)=E_{\rm mode}(\mu^2)+ \frac{1}{2}\left(\psi(1)-\psi(-1/2)\right)\frac{\xi C_2(\xi)}{4\pi^2},
\label{eff-modes}
\end{equation} 
where $\psi(s)$ is the digamma function and $C_2$ is the second Seeley-De Witt coefficient.

While the $C_2$ coefficient for a geometry consisting of two parallel plates vanishes, in the case of a spherical bag it is nonzero. As noted in ref. \cite{wipf}, this difference reflects the inherent ambiguity introduced in the Casimir energy when removing the polar 
part in eq. (\ref{zetatres}), by choosing the principal value, such as in eq. (\ref{modezeta}), or by using other possible prescriptions.
 
According to ref. \cite{wipf}, $C_2$ can be read from the variation of $E_{\rm mode}$
under a change of the parameter $\xi$, 
\begin{equation}
E_{\rm mode}|_{\xi'}-E_{\rm mode}|_{\xi}=- \frac{\xi C_2(\xi)}{4\pi^2}\ln \frac{\xi'}{\xi}.
\label{varxi}
\end{equation}
We will discuss below the particular form of $C_2$ in the bag model context, and we will see that both Casimir energy definitions, $E_C$ and $E_{\rm mode}$, lead in fact to the same physical implications.

It is important to remark that in the superposition we have derived for the Casimir energy
$E_C$ (cf. eq. (\ref{summodes2})), the Casimir energies $E_C(\mu^2)$ encode the geometry, while the factor $\beta(\mu^2)$ encodes the analytic structure of the model. 
Note also that the propagator for a free massive field of mass $m$ would give $\beta(\mu^2)=\delta(\mu^2-m^2)$ in eq. (\ref{dec}), and eq. (\ref{summodes2})
would give $E_C=E_C(m^2)$, as expected. 

In general, because of the presence of $E_C(\mu^2)$, our expression
corresponds to a physical situation where all the $\mu^2$ field modes
that participate in the representation (\ref{dec}) are confined,
that is, they see the boundary condition. From this point of view
our approach is quite natural, since confinement is supposed to
act over all the gluon modes represented by the model propagator
$G$. This should be contrasted with typical radiative QED corrections
to the Casimir effect where, in the computation strategy, the
boundary condition is seen by the free photon propagator but not
by the continuum of electron-positron modes represented in the
vacuum polarization \cite{bor1}. In these cases the overall result
is a renormalization of the sphere radius, the distance between
plates, etc. This renormalization can be traced back to the effect
of electron-positron pairs, which are not confined by the boundary
conditions, thus leading to an effective enlargement of the
confining region.

We also note that since our formula relies on the superposition of the
usual Casimir energies, obtained from the regularized partition function
for field modes with mass parameter $\mu^2$, the form of the divergences we will obtain here are the standard ones. For a detailed discussion see refs. \cite{romeo,bordagrep,bordag}. These divergent terms will renormalize similar terms in the bag model to be fixed by
phenomenology. For instance, the divergent volume term will
renormalize the bag constant $B$.

Now, according to our discussion at the end of section \S \ref{irmodel},
let us consider
for the model gluon propagator an ansatz of the form given in eq. (\ref{SD}).
In this case, eq. (\ref{discontinuity}) leads to,
\begin{eqnarray}
\beta(k^2)&=&\frac{1}{2\pi i}\delta \left[ \frac{\lambda}{k^2}-
\frac{1+\lambda}{k^2-\Lambda^2}
+\frac{d\ln {\cal R}(\cal A)}{d\cal A}\right]
\nonumber \\
&=& -\lambda \delta(k^2)+(1+\lambda)\delta(k^2-\Lambda^2)+ \beta_{\cal R}(k^2),
\label{beta}\nonumber \\
\end{eqnarray}
where we have used eq. (\ref{definit}), the property
\[
1/(k^2+i\epsilon)-1/(k^2-i\epsilon)=-2\pi i\, \delta(k^2),
\]
and the definition,
\begin{equation}\label{betar}
\beta_{\cal R}(k^2)=\frac{1}{2\pi i}\delta \left[\frac{d\ln {\cal R}(\cal
A)}{d\cal A}\right].
\label{betaRR}
\end{equation}
Then, from eqs. (\ref{summodes2}) and (\ref{beta}), the Casimir energy for
the infrared modified model
results,
\begin{equation}
E_C=-\lambda E_C(0)+ (1+\lambda) E_C(\Lambda^2) + \int_0^\infty
d\mu^2\beta_{\cal R}(\mu^2) E_C (\mu^2).
\label{result}
\end{equation}
Of course, if a free massless propagator were considered, that is,
$\lambda=-1$ and ${\cal R}\equiv 1$ (cf. eq. (\ref{SD})),
we would have $\beta(\mu^2)=\delta(\mu^2)$, and the Casimir energy
$E_C=E_C(0)$ for a massless field would be reobtained. 

In that case, the zeta function calculation for the regularized 
energy mode summation in eq. (\ref{modezeta}) has been performed in ref. \cite{romeo}, obtaining,
\begin{equation}
E_{\rm mode}(0)= \frac{1}{R}\left[ 0.08392 +\frac{8}{315\pi} \ln (\xi R)\right],
\label{emodes0}
\end{equation}
which corresponds to photon modes confined inside the bag. Using this expression in 
eq. (\ref{varxi}), we read,
\begin{equation}
\frac{\xi C_2(\xi)}{4\pi^2}=-\frac{8}{315\pi}\frac{1}{R},
\end{equation}
which corresponds to an additional $1/R$ term when passing from the Casimir energy
defined by $E_{\rm mode}(0)$ to $E_C(0)$ in eq. (\ref{eff-modes}); evaluating the digamma functions, we get,
\begin{equation}
E_C(0)= \frac{1}{R}\left[ 0.08640 +\frac{8}{315\pi} \ln (\xi R)\right].
\label{eeff0}
\end{equation}
In ref. \cite{romeo}, the zeta function result for the summation of energy modes
in eq. (\ref{emodes0}) has been compared with that obtained in ref. \cite{milton3},
\begin{equation}
\frac{1}{R}\left[ 0.08984 +\frac{8}{315\pi} \ln (\delta/8)\right],
\label{egreen}
\end{equation}
found by the Green's function method, where $\delta$ is a cut-off associated with a nonzero 
``skin depth'' representing a realistic boundary, instead of a sharp mathematical one.
As noted in that reference, the $1/R$ parts obtained from the first term in eqs.
(\ref{emodes0}) and (\ref{egreen}), differ within a $6.7\%$, and in fact, there is no reason why these parts should be equal, as they may vary by just changing the values of the parameters $\delta$ and $\xi$. The same situation applies to the $1/R$ part coming from $E_C$, obtained from the first term in eq. (\ref{eeff0}).

In a bag model context, the point is that a natural scale exists which permits to derive
some qualitative and semiquantitative conclusions (see ref. \cite{Milton} and references therein).

For instance, the plausible value for $\ln (\delta/8)$ is of order one, when realistic boundary conditions with a ``skin depth'' of about $10\%$ of a typical bag radius is considered, see ref. \cite{Milton}. There, a first crude estimate $\sim +0.7/R$ has been obtained for the zero point energy by dropping the logarithm term in eq. (\ref{egreen}) (which corresponds to $10\%$ of the first term), and multiplying by a factor eight counting for the number of gluons. As stated in \cite{Milton}, it is certainly very hard to doubt about the sign of the effect. 

In the zeta function regularization scenario, according to refs. \cite{romeo}, \cite{Milton} and \cite{bordagrep}, in a pure QCD context $\xi$ should be associated
with the energy scale parameter $\Lambda_{QCD}$; as a consequence, we will see that the discussion in the above paragraph is also realized in this type of scenario.

Let us consider $\xi=\Lambda=600\,{\rm Mev}$ (a typical value for the fitting of the nonperturbative propagator (\ref{SD})), and the radius $R$ taking values around $R_0=\frac{1}{200}\, {\rm Mev}^{-1}$ 
(a typical value for a hadron radius). We see that for this choice the logarithm in eqs. (\ref{emodes0}) and (\ref{eeff0}) is of order one, and again the corresponding second 
term represents about $10\%$ of the first term. In general, we can write eq. (\ref{eeff0}) as,
\begin{equation}
E_C(0)= \frac{1}{R}\left[ 0.08640 +\frac{8}{315\pi} \ln (\Lambda R_0)+
\frac{8}{315\pi} \ln (R/R_0)\right].
\label{eeffC}
\end{equation}
After multiplying by eight colors, the first two terms in the bracket give a contribution $+0.75/R$. Then, we see that in the zeta function regularization scenario, a clear scale for the Casimir effect again arises. The Casimir energy in the ``photon-like'' bag context can be considered as consisting of a $\sim +0.7/R$ part, plus logarithmic corrections.

We also note that even in an $R$ interval that covers two orders of magnitude around $R_0$, from $(1/10)\,R_0$ to $10\, R_0$, the whole bracket in eq. (\ref{eeffC}) remains of order one, varying from $+0.6$ to $+0.9$. Then, in this scenario it is also very hard to doubt about the repulsive character of the effect. For ``photon-like'' gluon modes confined inside the bag, in order to change the sign of the Casimir energy and comply with phenomenology, the typical hadron radius $R_0$ and typical scale $\Lambda$ in pure QCD should lead to a negative contribution of order one coming from the first two terms in eq. (\ref{eeffC}). This would happen for $\Lambda R_0 < 10^{-9}$, which is an unrealistic situation.

Coming back to the infrared modified bag model, we observe that the
first term in eq. (\ref{result}) only involves the soft $\mu^2=0$
modes. On the other hand, note that the second term in eq.
(\ref{result}) is associated with a massive Casimir effect 
$E_C(\Lambda^2)$ which is expected to be suppressed. 

This can be seen by considering the general form of the Casimir effect for a massive scalar field. In ref. \cite{bordag}, the zeta function regularized expression
for the Casimir energy has been computed; the associated divergent terms and ambiguities 
always display nonnegative powers of the mass of the field. On the other hand, as discussed in ref. \cite{bordagrep}, in the massive case there is a natural renormalization prescription, requiring that in the infinite mass limit all effects coming from quantum fluctuations should vanish. In this manner, after renormalization, it is obtained,
\begin{equation}
E_C(\Lambda^2)=\frac{f(\Lambda R)}{R},
\label{ECadim}
\end{equation}
where $f$ is a well defined, finite and unambiguous function. 

Although a simple analytic expression for $f(\Lambda R)$ is lacking, its numerical evaluation has been performed in ref. \cite{bordag}, and the result for the interior problem with Dirichlet boundary conditions has been presented by plotting $f=R E_C$ as a function of the dimensionless variable $\Lambda R$.
By using this information, it is simple to estimate the second term in eq. (\ref{result}),
taking as before $\Lambda=600\, {\rm Mev}$, and $R=\frac{2}{3} \frac{1}{200}$ or $R=\frac{1}{200}$ (typical values in ${\rm Mev}^{-1}$ for the pion and proton radius, respectively). Around these values, we have $E_C(\Lambda^2)=f(2)/R$ or $f(3)/R$, respectively. From ref. \cite{bordag}, we can extract the estimate $f(2)\approx .0002$ which, after including a $2\times 8$ factor 
counting the number of physical polarizations times the number of colors, implies that the second term in eq. (\ref{result}) typically corresponds to a $1 \%$ of the first term ($\lambda$ is of order 1). The analysis for $f(3)$ gives an even higher suppression than the previous one.

Note that in our model the first two factors in eq. (\ref{SD}) are the relevant ones to interpolate between an ultraviolet asymptotically free and an infrared vanishing behavior of the effective gluon propagator. So that, in principle, the model could be defined with
${\cal R}\equiv 1$ and the third term in eq. (\ref{result}) would be absent. 

However, it is interesting to analyze the effect of a nontrivial ${\cal R}$ factor. 
To be more specific, let us consider for instance the ansatz for ${\cal R}=\alpha^{-\gamma}$
(cf. eq. \ref{Rfact}), now extended over the whole range of momenta according to \cite{Alkofer04},
\begin{equation}
{\cal R}= ({\cal A}-\Lambda^2)^{\gamma}\alpha'^{-\gamma},
\label{rfact}
\end{equation}
\begin{equation}
\alpha'=-\alpha(0)\Lambda^2+\frac{4\pi}{b}{\cal A}\left( \frac{1}{\ln
(-{\cal A}/\Lambda^2)}+\frac{\Lambda^2}{{\cal A}+\Lambda^2}\right),
\end{equation}
where $\alpha(0)=8.915/N_c$, $\gamma=(-13N_c + 4 N_f)/(22N_c-4N_f)$, $b=(11N_c-2N_f)/3$, $N_c$ and $N_f$ being the number of colors and flavors, respectively, and the $i\epsilon$ prescription in the variable ${\cal A}=k^2$ is understood. This form implies a cut in Minkowski space along
the timelike $k^2$ axis with branch point at $k^2=0$, and the contribution
to the discontinuity $\beta_{\cal R}(\mu^2)$ defined in eq. (\ref{betar}) has support starting at $\mu^2=0$. 

Using eq. (\ref{betaRR}), we have, 
\begin{eqnarray}
\lefteqn{\beta_{\cal R}(k^2)=-\gamma \delta(k^2-\Lambda^2)+\beta'_{\cal R}(k^2),}
\nonumber \\
&&\beta'_{\cal R}(k^2)=-\frac{\gamma}{2\pi i}\, \delta \left[ \frac{d~}{d{\cal A}} \ln \alpha' \right].
\end{eqnarray}
The first term in $\beta_{{\cal R}}$ simply renormalizes the second term in eq. (\ref{result}), that is,
\begin{eqnarray}
\lefteqn{E_C=-\lambda E_C(0)+}\nonumber \\
&&+ (1+\lambda-\gamma) E_C(\Lambda^2) + \int_0^\infty d\mu^2\beta'_{\cal R}(\mu^2) E_C (\mu^2),\nonumber \\
\label{resultf}
\end{eqnarray}
and as $\gamma$ is of order one, the same analysis we have done before can be applied to conclude that the second term in eq. (\ref{resultf}) is suppressed.

\begin{figure}
%\vspace{1cm}
%\noindent
%\hspace{-2.0 cm}
\epsfxsize=8 cm
\epsfysize=7 cm
\epsfbox{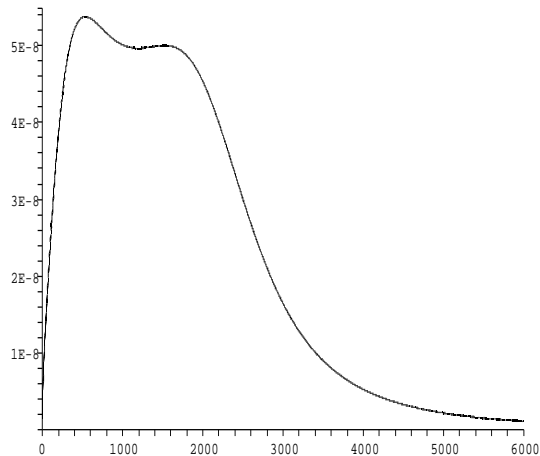}
%\vspace{1 cm}
%\hspace{3.0 cm}
%\epsfxsize=6 cm
%\epsfysize=6 cm
%\epsfbox{nn70.eps}

\caption{ This is a plot of $\beta'_{\cal R}$ (in ${\rm Mev}^{-1}$), given in eq. (\ref{cform}), vs. the mass parameter $\mu$ (in Mev). We have considered $N_c=3$, and $N_f=0$ (quenched case).}
\label{fig1}
\end{figure}

On the other hand, $\beta'_{{\cal R}}$ can be written in terms of the real and imaginary parts of $\alpha'({\cal A}+i\epsilon)$,
\begin{equation}
\beta'_{\cal R}(k^2)=-\frac{\gamma}{\pi}\, \frac{d~}{d{\cal A}} \arctan \left(\frac{\Re (\alpha')}{\Im (\alpha')}\right),
\label{cform}
\end{equation} 
\begin{equation}
\Re (\alpha')= -\alpha(0)\Lambda^2+\frac{4\pi}{b}{\cal A}
\left( \frac{{\ln (\cal A}/\Lambda^2)}{\ln^2({\cal A}/\Lambda^2)+\pi^2}+
\frac{\Lambda^2}{{\cal A}+\Lambda^2}\right),
\end{equation}
\begin{equation}
\Im (\alpha')=\frac{4\pi^2}{b}\frac{{\cal A}}{\ln^2({\cal A}/\Lambda^2)+\pi^2}.
\end{equation}
In Fig. \ref{fig1}, we plot $\beta'_{{\cal R}}(\mu^2)$ as a function of $\mu$. We observe that the function $\beta'_{{\cal R}}$ displays a definite sign, weighting the massive Casimir energies in the third term of eq. (\ref{resultf}).

The total weight that is distributed over the whole range of masses is,
\begin{equation}
\int_0^\infty
d\mu^2\beta'_{\cal R}(\mu^2)=-\frac{\gamma}{\pi}\, 
\left. \arctan \left(\frac{\Re (\alpha')}{\Im (\alpha')}\right)\right|_0^{\infty}=-\gamma,
\label{totw} 
\end{equation}
which can also be verified numerically; for $N_c=3$, and $N_f=0$ (quenched theory), 
we have $-\gamma\approx 0.59$.

Now, we can use the general parametrization of the Casimir energy in the massive case, given in eq. (\ref{ECadim}), with $\Lambda \rightarrow \mu$, to write,  
\begin{equation}
\int_0^\infty
d\mu^2\beta'_{\cal R}(\mu^2) E_C (\mu^2)=\frac{1}{R}
\int_0^\infty d\mu^2\beta'_{\cal R}(\mu^2) f(\mu R).
\label{betacomf}
\end{equation}
In order to obtain a bound for the contribution of this term, we can consider, as before, a typical radius $R=\frac{1}{200}\, {\rm Mev}^{-1}$, and divide the $\mu^2$ integration in three intervals $I_i$, $i=0,1,2$: from $0$ to $(200~{\rm Mev})^2$, $(200~{\rm Mev})^2$ to $(400~{\rm Mev})^2$ and $(400~{\rm Mev})^2$ to $\infty$.

Note also in Fig. \ref{fig1} that in the limit $\mu^2 \to 0$, we have 
$\beta'_{{\cal R}}\to 0$. Indeed, it can be verified that $\beta'_{{\cal R}}$ tends to zero like $\sim 1/(\ln \mu^2)^2$. Therefore, in eq. (\ref{betacomf}), the contribution coming from a region very close to $\mu^2=0$ is suppressed. Outside from this small region, it has been shown in ref. \cite{bordag} that $f(\mu R)$ is positive definite, displaying a maximum value $f_0\approx .0030$ in the region $I_0$. In the region $\mu R \geq 1$, the function $f(\mu R)$ decreases  monotonically to zero. Then, the maximum values in regions $I_1$ and $I_2$ are given by $f_1=f(1)\approx .0005$ and $f_2=f(2)\approx .0002$, respectively. These values have been estimated from the numerical result presented in that reference.

Then, replacing  in each interval $f(\mu R)$ by $f_i$, we obtain,
\begin{equation}
\int_0^\infty d\mu^2\beta'_{\cal R}(\mu^2) f(\mu R)\leq 
\sum_{i=0}^2 f_i \int_{I_i} d\mu^2\beta'_{\cal R}(\mu^2)
\label{cota} 
\end{equation}

With regard to the integrals in eq. (\ref{cota}), we obtained $.001$ and $.006$ in the intervals $I_0$ and $I_1$, while in the interval $I_2$ the integral completes the total weight $.59$. In this manner, around a typical radius of $R=\frac{1}{200}$ Mev$^{-1}$,    
\begin{equation}
\int_0^\infty
d\mu^2\beta'_{\cal R}(\mu^2) E_C (\mu^2)\leq 10^{-4}/R.
\label{inequality}
\end{equation}
After including  a factor for the total number of polarizations, we see that this term represents a contribution less than $0.3 \%$ when compared with the first term in eq. (\ref{resultf}). 

As expected, the effect of the boundaries on the finite part of the zero point energy
is dominated by the soft modes $\mu^2=0$ present in the first term of eq. (\ref{resultf}).
We have also considered a number of flavors $N_f=3$, obtaining no essential modifications 
in the analysis. 

Then, in the context of the infrared modified bag model, the $1/R$ part of the renormalized vacuum energy turns out to be $-Z/R$, with $Z\sim +0.7 \lambda$, which is attractive for an infrared vanishing model propagator ($\lambda > 0$). 

\section{Conclusions}

In this article, we have computed the Casimir energy in a bag model
containing a modified gluon propagator.
Our motivation comes from previous attempts to describe the vacuum energy in the
bag in terms of Casimir energy calculations. These attempts, based in a model
where the gluons are associated with massless fields confined on a spherical
region, failed to describe the attractive nature of the vacuum energy
necessary to adjust hadron spectroscopy.

In the modified framework we have considered, we were able to
clearly trace the effects introduced by the analytic structure of
the model, on the one hand, and the geometry of the boundaries, on
the other (see equation (\ref{summodes2})). In this context, we can see
that the introduction of perturbative information in the above mentioned  
massless gluon models is not enough to render the finite part of the zero point
energy attractive; if only a running coupling were taken into account, the
effect would be suppressed when compared with the main Casimir effect coming from
the pole at $k^2=0$, which is repulsive.

Then, in order to understand the attractive nature of the
Casimir energy due to gluons, the consideration of intrinsic nonperturbative
effects is fundamental, however, an analytic approach from first principles 
is a formidable challenge. 
For this reason, we have introduced the
simple infrared modified bag model with an effective gluon propagator.
Within this context, we have seen that the repulsive or attractive
nature of the vacuum energy depends on whether the model
propagator describes ``confined'' gluons with infrared vanishing
behavior, like in the Schwinger-Dyson and Gribov scenarios, or it
is singular at $k^2=0$. This last situation is verified, for
instance, when a free $1/k^2$ or a ``confining'' $1/k^4$ model
gluon propagator is used.

From this point of view, we see that an infrared vanishing  gluon propagator
is preferred as it corresponds to an attractive term $\sim -0.7\lambda/R$, $\lambda>0$, in the renormalized vacuum energy. Hadron phenomenology requires a gluon contribution
of the form $-Z/R$, with $Z$ of order one (see the discussion in ref.
\cite{DJ}). In the simple model we have presented,
this would be achieved with an infrared behavior of the model gluon propagator
of the form $\sim (\bar{k}^2)^\lambda$, with $\lambda$ of order one, which corresponds 
to a typical order of magnitude appearing in those pure QCD scenarios where an infrared vanishing propagator for ``confined'' gluons is obtained.

As further studies suggested by the present work, we point out the consideration of similar  nonperturbative effects in the fermion sector, as well as in other bag models where the hyperfine hadron structure is analyzed \cite{W}. 

\begin{acknowledgments}
The Conselho Nacional de Desenvolvimento Cient\'{\i}fico e
Tecnol\'{o}gico (CNPq) and the Funda{\c {c}}{\~{a}}o de Amparo
{\`{a}} Pesquisa do Estado do Rio de Janeiro (FAPERJ) are acknowledged for the financial support.
\end{acknowledgments}

\end{document}